\documentclass[twoside,twocolumn,9pt]{article}
\usepackage{extsizes}
\usepackage[super,sort&compress,comma]{natbib} 
\usepackage[version=3]{mhchem}
\usepackage[left=1.5cm, right=1.5cm, top=1.785cm, bottom=2.0cm]{geometry}
\usepackage{balance}
\usepackage{mathptmx}
\usepackage{sectsty}
\usepackage{graphicx} 
\usepackage{lastpage}
\usepackage[format=plain,justification=justified,singlelinecheck=false,font={stretch=1.125,small,sf},labelfont=bf,labelsep=space]{caption}
\usepackage{float}
\usepackage{fancyhdr}
\usepackage{fnpos}
\usepackage[english]{babel}
\addto{\captionsenglish}{%
  
}
\usepackage{array}
\usepackage{droidsans}
\usepackage{charter}
\usepackage[T1]{fontenc}
\usepackage[usenames,dvipsnames]{xcolor}
\usepackage{setspace}
\usepackage[compact]{titlesec}
\usepackage{hyperref}

\usepackage{epstopdf}%This line makes .eps figures into .pdf - please comment out if not required.

\begin{document}

\pagestyle{fancy}
\thispagestyle{plain}
\fancypagestyle{plain}{
%%%HEADER%%%
\renewcommand{\headrulewidth}{0pt}
}
%%%END OF HEADER%%%

%%%PAGE SETUP - Please do not change any commands within this section%%%
\makeFNbottom
\makeatletter
\renewcommand\LARGE{\@setfontsize\LARGE{15pt}{17}}
\renewcommand\Large{\@setfontsize\Large{12pt}{14}}
\renewcommand\large{\@setfontsize\large{10pt}{12}}
\renewcommand\footnotesize{\@setfontsize\footnotesize{7pt}{10}}
\makeatother

\renewcommand{\thefootnote}{\fnsymbol{footnote}}
\renewcommand\footnoterule{\vspace*{1pt}% 
\color{black}\vspace*{5pt}} 
\setcounter{secnumdepth}{5}

\makeatletter 
\renewcommand\@biblabel[1]{#1}            
\renewcommand\@makefntext[1]% 
{\noindent\makebox[0pt][r]{\@thefnmark\,}#1}
\makeatother 
\renewcommand{\figurename}{\small{Fig.}~}
\sectionfont{\sffamily\Large}
\subsectionfont{\normalsize}
\subsubsectionfont{\bf}
\setstretch{1.125} %In particular, please do not alter this line.
\setlength{\skip\footins}{0.8cm}
\setlength{\footnotesep}{0.25cm}
\setlength{\jot}{10pt}
\titlespacing*{\section}{0pt}{4pt}{4pt}
\titlespacing*{\subsection}{0pt}{15pt}{1pt}
%%%END OF PAGE SETUP%%%

%%%FOOTER%%%
\fancyfoot{}
\fancyfoot[LO,RE]{\vspace{-7.1pt}}
\fancyfoot[CO]{\vspace{-7.1pt}\hspace{13.2cm}}
\fancyfoot[CE]{\vspace{-7.2pt}\hspace{-14.2cm}}
\fancyfoot[RO]{\footnotesize{{\hspace{2pt}\thepage}}}
\fancyfoot[LE]{\footnotesize{{\thepage\hspace{3.45cm}}}}
\fancyhead{}
\renewcommand{\headrulewidth}{0pt} 
\renewcommand{\footrulewidth}{0pt}
\setlength{\arrayrulewidth}{1pt}
\setlength{\columnsep}{6.5mm}
\setlength\bibsep{1pt}
%%%END OF FOOTER%%%

%%%FIGURE SETUP - please do not change any commands within this section%%%
\makeatletter 
\newlength{\figrulesep} 
\setlength{\figrulesep}{0.5\textfloatsep} 

\newcommand{\topfigrule}{\vspace*{-1pt}% 
\noindent{\color{black}\rule[-\figrulesep]{\columnwidth}{1.5pt}} }

\newcommand{\botfigrule}{\vspace*{-2pt}% 
\noindent{\color{black}\rule[\figrulesep]{\columnwidth}{1.5pt}} }

\newcommand{\dblfigrule}{\vspace*{-1pt}% 
\noindent{\color{black}\rule[-\figrulesep]{\textwidth}{1.5pt}} }

\makeatother
%%%END OF FIGURE SETUP%%%

%%%TITLE, AUTHORS AND ABSTRACT%%%
\twocolumn[
\noindent\LARGE{\textbf{Atomistic insights into hydrogen migration in IGZO from machine-learning interatomic potential: linking atomic diffusion to device performance}}
\vspace{0.3cm} \\

\noindent\large{Hyunsung Cho,\textit{$^{a}$} Minseok Moon,\textit{$^{a}$} Jaehoon Kim,\textit{$^{a}$} Eunkyung Koh,\textit{$^{b}$} Hyeon-Deuk Kim,\textit{$^{b}$} Rokyeon Kim,\textit{$^{b}$} Gyehyun Park,\textit{$^{b}$} Seungwu Han,\textit{$^{ac}$} and Youngho Kang$^{\ast}$\textit{$^{d}$}} \\

\noindent\normalsize{Understanding hydrogen diffusion is critical for improving the reliability and performance of oxide thin-film transistors (TFTs), where hydrogen plays a key role in carrier modulation and bias instability. In this work, we investigate hydrogen diffusion in amorphous IGZO ($a$-IGZO) and $c$-axis aligned crystalline IGZO (CAAC-IGZO) using machine learning interatomic potential molecular dynamics (MLIP-MD) simulations. We construct accurate phase-specific MLIPs by fine-tuning SevenNet-0, a universal pretrained MLIP, and validate the models against a comprehensive dataset covering hydrogen-related configurations and diffusion environments. Hydrogen diffusivity is evaluated over 650–1700~K, revealing enhanced mobility above 750~K in $a$-IGZO due to the glassy matrix, while diffusion at lower temperatures is constrained by the rigid network. Arrhenius extrapolation of the diffusivity indicates that hydrogen in $a$-IGZO can reach the channel/insulator interface within $10^4$ seconds at 300–400~K, likely contributing to negative bias stress-induced device degradation. Trajectory analysis reveals that long-range diffusion in $a$-IGZO is enabled by a combination of hydrogen hopping and flipping mechanisms. In CAAC-IGZO, hydrogen exhibits high in-plane diffusivity but severely restricted out-of-plane transport due to a high energy barrier along the $c$-axis. This limited vertical diffusion in CAAC-IGZO suggests minimal impact on bias instability. This work bridges the atomic-level hydrogen transport mechanism and device-level performance in oxide TFTs by leveraging large-scale MLIP-MD simulations.
}
 \vspace{0.6cm}
  ]
%%%END OF TITLE, AUTHORS AND ABSTRACT%%%

%%%FONT SETUP - please do not change any commands within this section
\renewcommand*\rmdefault{bch}\normalfont\upshape
\rmfamily
\section*{}
\vspace{-1cm}

%%%FOOTNOTES%%%
\footnotetext{\textit{$^{a}$~Department of Materials Science and Engineering and Research Institute of Advanced Materials, Seoul National University, Seoul 08826, Republic of Korea.}}
\footnotetext{\textit{$^{b}$~Display Research Center, Samsung Display Company Ltd.,} Yongin-si 17113, Republic of Korea.}
\footnotetext{\textit{$^{c}$~Korea Institute for Advanced Study, Seoul 02455, Republic of Korea.}}
\footnotetext{\textit{$^{d}$~Department of Materials Science and Engineering, Incheon National University, Incheon 22012, Republic of Korea. E-mail: youngho84@inu.ac.kr}}

%%%END OF FOOTNOTES%%%

%%%MAIN TEXT%%%%
\section{Introduction}
Over the past decade, amorphous indium gallium zinc oxide ($a$-IGZO) and related oxides have remained at the forefront of metal oxide electronics, serving as a key channel material for thin-film transistors (TFTs) in display technologies~\cite{nomura_2004, display_kwon_2008, review_kamiya_2010, review_ide_2019}. This prominence is attributed to its numerous advantages, including compatibility with low-temperature processing ($<200^{\circ}\mathrm{C}$), relatively high electron mobility ($>10~\mathrm{cm}^2/\mathrm{Vs}$) compared to amorphous silicon ($<1~\mathrm{cm}^2/\mathrm{Vs}$), a large band gap over 3~eV that enables low power consumption and high optical transparency, a low threshold voltage ($<1~\mathrm{V}$), and a small subthreshold swing ($<0.1~\mathrm{V/dec}$)~\cite{nomura_2004, review_kamiya_2010,review_ide_2019,ss_samanta_2020}. The success of oxide TFTs in displays has recently expanded the application of $a$-IGZO to emerging fields such as next-generation computing and artificial intelligence (AI), including neuromorphic transistors, offering a pathway to overcome the limitations of conventional Si-based complementary metal–oxide–semiconductor (CMOS) technology~\cite{cmos_lopez_2024}. Additionally, due to its back-end-of-line (BEOL) compatibility~\cite{beol_li_2022}, the $a$-IGZO channel has attracted growing interest for use in monolithic three-dimensional (M3D) semiconductor devices~\cite{m3d_an_2022}.

Hydrogen is a ubiquitous element and is known to be readily incorporated into the oxide channel during the fabrication of oxide TFTs. Two primary pathways have been suggested for hydrogen incorporation into the $a$-IGZO channel. First, hydrogen can be introduced during channel formation via sputtering or pulsed laser deposition (PLD), mainly due to residual hydrogen-containing species in the reaction chamber, such as H$_2$O and H$_2$~\cite{h_diffusivity_nomura_2013,h_sputtering_tang_2015,review_ide_2019}. 
The hydrogen concentration in the channel is subsequently determined by the extent of hydrogen out-diffusion during post-deposition annealing at elevated temperatures. Second, hydrogen can also be introduced during the deposition of gate dielectric or passivation over-layers, such as SiO$_2$ or Al$_2$O$_3$, using plasma-enhanced chemical vapor deposition (PECVD) or atomic layer deposition(ALD)~\cite{h_passivation_instability_toda_2014, h_diffusion_effect_nam_2018, h_diffusion_noh_2022}. In this case, hydrogen atoms diffuse from the overlying films into the underlying $a$-IGZO layer during deposition and subsequent annealing processes.

Once incorporated, hydrogen acts as a critical defect that significantly influences the electrical properties of $a$-IGZO and, consequently, the performance of oxide TFTs based on it. According to previous studies, the hydrogen concentration must be carefully controlled to achieve an appropriate threshold voltage ($V_\mathrm{{th}}$) and subthreshold swing ($SS$)~\cite{h_nbis_kim_2013}. Particularly, excessive hydrogen incorporation can result in a highly negative $V_\mathrm{{th}}$ and a large $SS$~\cite{nbs_h_oh_2013}. This degradation occurs because hydrogen acts as shallow $n$-type donors in $a$-IGZO~\cite{vo_h_kamiya_2010}, leading to a significant increase in free electron concentration—often exceeding 10$^{19}$~cm$^{-3}$—when present in large amounts~\cite{h_anneal_ne_kamiya_2009}. 

Furthermore, hydrogen impurities have been identified as a key source of electrical instabilities in oxide TFTs under gate bias and illumination stress conditions, such as negative bias stress (NBS), negative bias illumination stress (NBIS), and positive bias stress (PBS)~\cite{h_nbis_kim_2013, hi_ho_nbis_noh_2013, h_instability_kim_2020, h_instability_chen_2020, h_passivation_instability_toda_2014, h_conc_pbts_jeon_2017, defects_meux_2018}. In particular, under NB(I)S conditions, positively charged hydrogen species are supposed to diffuse from the channel toward the gate insulator interface. This diffusion leads to the accumulation of positive charge at the interface, resulting in a negative shift in $V_\mathrm{th}$ and subsequent degradation in device performance.

The foregoing discussion underscores that hydrogen diffusion in $a$-IGZO plays a pivotal role in determining the electrical performance and bias-related instability of oxide TFTs. Therefore, a thorough understanding of its behavior is essential for the development of high-performance and reliable oxide TFTs. Despite its significance, little is known about the detailed behavior of hydrogen diffusion in $a$-IGZO. In particular, hydrogen is difficult to detect using conventional experimental techniques due to its exceptionally low atomic mass~\cite{h_detection_shvachko_1998,h_review_chen_2025}. Consequently, analyzing the hydrogen diffusion process—which requires precise, time- and space-resolved quantification of hydrogen concentration—remains highly challenging and is often accompanied by substantial uncertainty. Accordingly, only a limited number of experiments have reported quantitative analysis on hydrogen diffusion in $a$-IGZO or similar oxide semiconductors~\cite{h_diffusivity_nomura_2013,in2o3_h_qin_2018,ga2o3_h_reinertsen_2020}.

To address this issue, several atomistic modeling studies have investigated hydrogen diffusion in $a$-IGZO by means of density functional theory (DFT) calculations. These studies computed activation energies ($E_{\rm{a}}$) along predefined hydrogen migration pathways~\cite{hi_ho_nbis_noh_2013} and performed on-lattice kinetic Monte Carlo (kMC) simulations to assess hydrogen dynamics~\cite{kMC_kim_2022}. However, the reliance on predefined migration paths may not capture the full diversity of diffusion mechanisms present in a highly disordered amorphous matrix. Furthermore, on-lattice kMC methods—where transition rates between artificial lattice sites are assigned based on a set of pre-calculated activation energies—neglect spatial and energetic correlations between successive diffusion events, potentially distorting the true kinetics of hydrogen transport. To gain physically meaningful insights and enable quantitative analysis of hydrogen diffusion, more direct approaches, such as molecular dynamics (MD) simulations, are required for modeling hydrogen migration processes in amorphous systems.

First-principles molecular dynamics (FPMD) simulations, based on DFT, have been widely adopted to investigate dynamical properties such as atomic diffusion~\cite{msd_he_2018,franckel_in2o3_2024}. Despite their high accuracy, the substantial computational cost of FPMD severely limits both the system size and simulation timescale, rendering it impractical for modeling long-range hydrogen migration in disordered materials like $a$-IGZO. As an alternative, machine learning interatomic potentials (MLIPs) have recently emerged as a powerful approach to overcome these limitations~\cite{mlff_unke_2021}. By learning energies and forces from DFT datasets, MLIPs achieve near-DFT accuracy while offering linear computational scaling ($\mathcal{O}(N)$), where $N$ is the number of atoms, making them suited for simulating large-scale amorphous systems over extended timescales. 

In this work, we present a theoretical study on hydrogen diffusion in $a$-IGZO and $c$-axis aligned crystalline IGZO (CAAC-IGZO) using MLIP-MD simulations. We first develop MLIPs suitable for modeling hydrogen migration in respective phases by fine-tuning SevenNet-0 (hereafter referred to as 7net-0), a universal pretrained MLIP model~\cite{sevennet_park_2024}. The fine-tuned MLIP models are thoroughly validated against a diverse dataset relevant to hydrogen-related configurations and diffusion environments in IGZO. We evaluate hydrogen diffusivity at elevated temperatures ranging from 650 to 1700 K by performing MD simulations using the fine-tuned MLIP models. The results reveal that hydrogen migration in $a$-IGZO is facilitated at temperatures above 750 K due to the glassy nature of the amorphous matrix, resulting in a lower activation energy. In contrast, at lower temperatures, hydrogen diffusion occurs through a more rigid atomic network, leading to a higher activation energy. Detailed analysis of the MD trajectories reveals that a combination of hydrogen hopping and flipping mechanisms enables long-range hydrogen diffusion in $a$-IGZO. In comparison, in CAAC-IGZO, hydrogen exhibits high diffusivity along the in-plane direction, while out-of-plane diffusion along the $c$-axis is significantly suppressed. Extrapolation of the diffusion coefficients to 300–400 K via the Arrhenius relation suggests that hydrogen in the $a$-IGZO channel can reach the channel/insulator interface within 10$^4$ seconds under NBS conditions, thereby accelerating device degradation. In contrast, negligible hydrogen diffusion in CAAC-IGZO along the $c$-axis suggests limited impact on bias-induced instability. Overall, by elucidating hydrogen migration in IGZO, our study provides key insights into its impact on the performance and electrical instability of oxide TFTs.

\section{Methods}

\subsection{Calculation setups}
As a MLIP, we employ 7net-0 (version July 11, 2024)~\cite{sevennet_park_2024}, a pretrained universal MLIP model trained on the MPtraj dataset~\cite{chgnet_deng_2023}. 7net-0 is built upon the NequIP architecture~\cite{nequip_batzner_2022}, which utilizes an equivariant graph neural network framework. The model has demonstrated high accuracy across a wide range of chemical systems, including oxides, and offers excellent scalability for large-scale molecular dynamics (MD) simulations on multi-GPU platforms—a key advantage over many other graph-based MLIP approaches.

7net-0 exhibits reasonable accuracy in predicting atomic structures and dynamics of IGZO in both amorphous and crystalline phases (Fig.~S1). However, in hydrogen-doped IGZO systems, we observe notable deficiencies in accuracy related to hydrogen-containing configurations (Fig.~S2). To address this, we fine-tune the pretrained 7net-0 model using a curated training set that includes hydrogen-related configurations with large prediction errors. Two fine-tuned MLIP models are developed: the $a$-FT model for $a$-IGZO and the $c$-FT model for CAAC-IGZO. The $a$-FT model is obtained by fine-tuning 7net-0 on a dataset containing hydrogen configurations in amorphous structures, while the $c$-FT model is trained using a dataset focused on hydrogen configurations in CAAC-IGZO (Table~S1). To preserve the model’s accuracy for defect-free systems, both fine-tuning datasets also include FPMD snapshots for corresponding IGZO systems without hydrogen, thereby mitigating the risk of catastrophic forgetting. Details of the fine-tuning procedure and the training set are provided in the Supporting Information. The training and validation root-mean-square errors (RMSEs) are reasonably small, comparable to previous MLIP studies~\cite{rmse_grunert_2025}, as summarized in Table~S2, illustrating the successful training of the fine-tuned models. 

MLIP-MD simulations are performed using the LAMMPS molecular dynamics package~\cite{lammps_thompson_2022} under the NVT ensemble using a Nos\'e--Hoover thermostat. A time step of 2 fs is used for systems without hydrogen, while it is reduced to 1 fs for hydrogen-containing systems to ensure MD stability. For MLIP-based structural relaxations and nudged elastic band (NEB) simulations, the Atomic Simulation Environment (ASE) interface is employed~\cite{ase_larsen_2017}. The convergence criterion for structural relaxation is set to 0.02~eV/\AA.

DFT calculations to generate the fine-tuning and test dataset are performed using the Vienna \textit{ab initio} Simulation Package (VASP)~\cite{vasp_kresse_1996}, with the Perdew--Burke--Ernzerhof (PBE) functional~\cite{pbe_perdew_1996} (version 52) as the exchange-correlation functional. A plane-wave cutoff energy of 520~eV is used for $a$-IGZO systems, while 500~eV for CAAC-IGZO systems. For structural relaxation, the convergence criterion is set to 0.02~eV/\AA. FPMD simulations are conducted under the NVT ensemble using a Nos\'e--Hoover thermostat. The Baldereschi $k$-point~\cite{baldereschi_1972} at (0.25, 0.25, 0.25) is used for Brillouin zone sampling in the FPMD simulations. To generate the fine-tuning dataset, we perform static DFT calculations on selected snapshots from FPMD simulations, using denser $k$-point grids of $3 \times 3 \times 3$ for $a$-IGZO and $2 \times 2 \times 2$ for CAAC-IGZO to obtain accurate atomic forces and total energies.

\subsection{Generation of $a$-IGZO structures}
Amorphous IGZO models, with a widely used stoichiometric ratio of In:Ga:Zn:O = 1:1:1:4 for decvie fabrication~\cite{in_cb_nomura_2007}, are generated using a melt-quench procedure~\cite{mq_kang_2014}. An initial atomic configuration is constructed in a cubic unit cell with a lattice parameter chosen to match the experimental density of 6.10~g/cm$^3$~\cite{density_hino_2012}. Initially, atoms are randomly distributed while avoiding unphysically short interatomic distances. The system undergoes a short 2-ps premelting step at 5000 K for initialization, followed by a 10-ps melting step at 2500 K. To generate diverse amorphous structures, additional MD simulations are performed at 2500 K, and snapshots are extracted every 2 ps. These configurations are subsequently quenched from 2500 K to 300 K at a cooling rate of 100 K/ps. Final amorphous structures are obtained through structural relaxation. Constructing the fine-tuning and test datasets from FPMD simulations is based on $a$-IGZO models containing 18 formula units (In$_{18}$Ga$_{18}$Zn$_{18}$O$_{72}$), with atoms added or removed as necessary to generate relevant defect configurations.

\subsection{Diffusivity calculations}

\begin{figure}
  \centering
  \includegraphics[width=0.8\columnwidth]{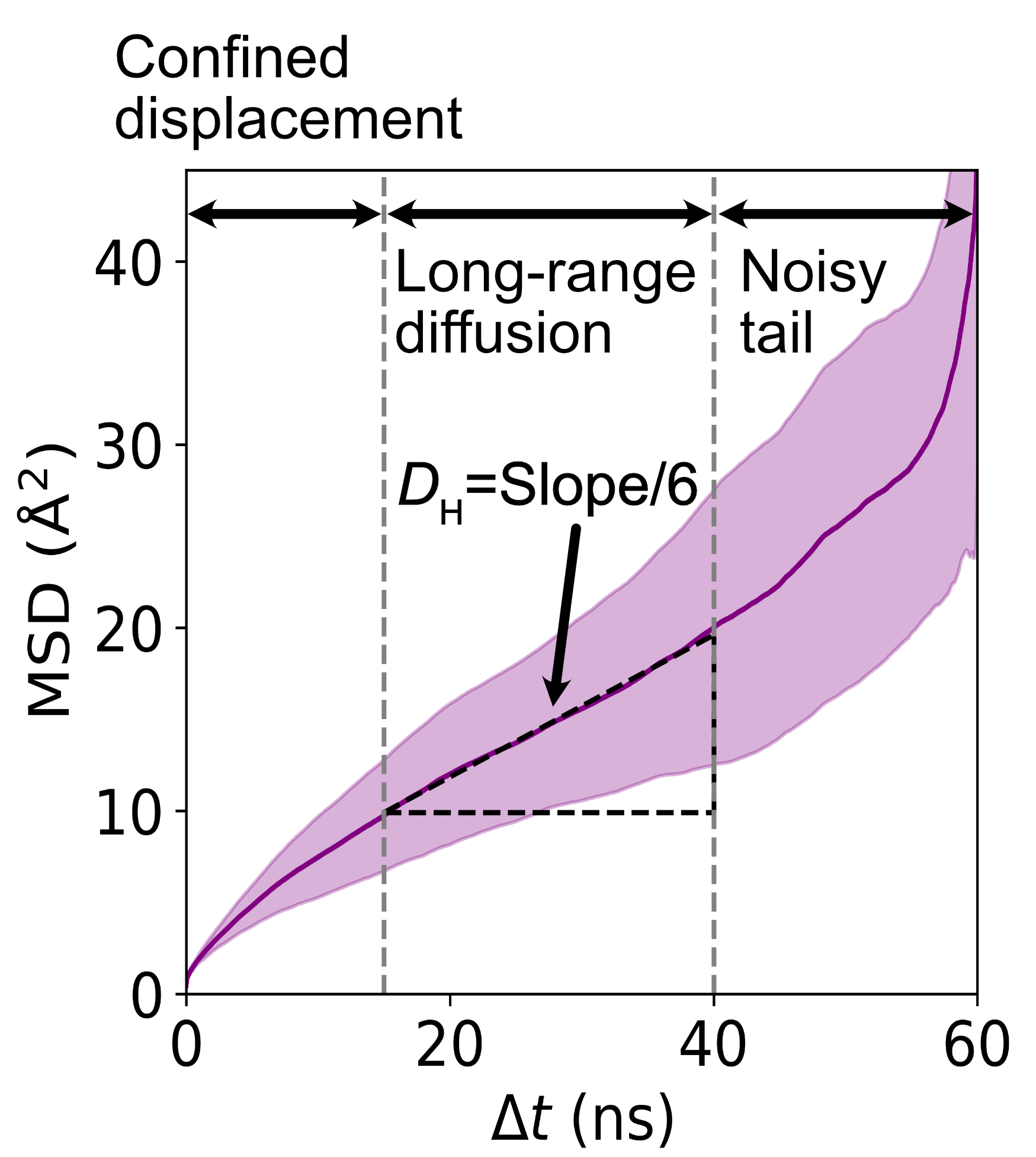}
  \caption{\small{Time evolution of the MSD of H atoms in $a$-IGZO at 650~$\mathrm{K}$. The MSD behavior depends on the displacement range, and diffusivity is extracted from the linear regime corresponding to long-range diffusion.}}
  \label{fig:msd_range}
\end{figure}

The self-diffusivity of hydrogen is evaluated from its mean square displacement (MSD) during MLIP-MD simulations. In this work, we compute the time-averaged MSD over an elapsed time $\Delta t$ from a given reference time $t$, defined as:
\begin{equation}
\label{eq:msd}
\mathrm{MSD}(\Delta t) = 
\frac{1}{N_{\mathrm{Cell}}} \sum_{e=1}^{N_{\mathrm{Cell}}}
\left[
\frac{1}{N_\mathrm{H}} \sum_{i=1}^{N_\mathrm{H}} 
\left( \frac{1}{N_{\Delta t}} \sum_{t=0}^{t_{\mathrm{tot}} - \Delta t} 
\left| \mathbf{r}_i^{(e)}(t + \Delta t) - \mathbf{r}_i^{(e)}(t) \right|^2 
\right)
\right]
\end{equation}
where $\mathbf{r}_i^{(e)}$ is the position of $i$-th hydrogen atom in the $e$-th supercell. $t_{\rm{tot}}$ denotes the total simulation time, and $N_{\rm{H}}$ is the number of hydrogen atoms in the supercell. To account for statistical variations, we perform an ensemble average over $N_{\rm{Cell}}$ independent supercell calculations. Subsequently, we evaluate H diffusivity ($D_{\rm{H}}$) from the slope of MSD-$\Delta t$ curve:
\begin{equation}
\label{eq:diff}
D_\mathrm{H} = \frac{\mathrm{MSD}(\Delta t)}{6\Delta t}.
\end{equation}
When calculating Eq.2, we use the portion of the MSD curve that exhibits clear linearity with respect to $\Delta t$. Specifically, as shown in Fig.~\ref{fig:msd_range}, we exclude the initial non-linear region (up to $\sim$10~\AA$^2$), which includes the ballistic regime, where vibrational motion is dominant, and hopping events confined to a small spatial region. On the other hand, as $\Delta t$ increases, the time-averaged MSD becomes less reliable, causing greater variance and deviation from the expected linear MSD behavior, due to insufficient statistical sampling~\cite{msd_he_2018}. Therefore, we extract the diffusivity from the intermediate region of the MSD data, which reflects statistically meaningful long-range hydrogen diffusion. The total simulation time $t_{\rm{tot}}$ is adjusted for each system to ensure an sufficiently long linear regime beyond $\rm{MSD}=10$ \AA$^2$ for reliable diffusivity estimation, with longer simulations conducted at lower temperatures to adequately capture hydrogen hopping events (Figs.~S10 and S11). The calculated temperature-dependent diffusivities are fitted to the Arrhenius relation: 
\begin{equation}
D_{\mathrm{H}}(T) = D_{0}\,\exp\!\left(-\frac{E_{\mathrm{a}}}{k_{\mathrm{B}}\,T}\right)
\label{eq:arrhenius_diffusion}
\end{equation}
where $D_{\rm{0}}$ is the pre-factor obtained from the Arrhenius fit,  $E_{\rm{a}}$ corresponds to the apparent activation energy, and $k_{\rm{B}}$ is the Boltzmann constant (Tables~S4–S7). 

For the amorphous phase, we use four ensemble simulation cells with the composition In$_{108}$Ga$_{108}$Zn$_{108}$O$_{432}$H$_{12}$, corresponding to $N_{\rm{H}} = 12$ and $N_{\rm{Cell}} = 4$. These structures are generated using MLIP-MD simulations via a melt-quench protocol and yield hydrogen concentrations of approximately $1.2 \times 10^{21}$cm$^{-3}$, which falls within the experimentally relevant range\cite{h_diffusivity_nomura_2013}. We observe that decreasing the hydrogen concentration has minimal impact on the calculated diffusivity (Table~S3), suggesting that hydrogen atoms are sufficiently separated and rarely interact with each other at the concentrations considered. This suggests that the resulting hydrogen diffusivity is applicable across a range of hydrogen concentrations observed in experiments. We also confirm that variations in oxygen stoichiometry—namely, oxygen-deficient and oxygen-rich environments—within the experimentally accessible range do not significantly affect hydrogen diffusivity. In the case of CAAC-IGZO, four ensemble simulation cells with the composition In$_{112}$Ga$_{112}$Zn$_{112}$O$_{448}$H$_{12}$ are constructed, yielding hydrogen concentrations comparable to those in the amorphous systems. In the crystalline structures, Ga and Zn atoms are randomly distributed over their respective cation sites, consistent with experimental observations~\cite{ga_zn_kimizuka_1995}.

\section{Results and discussion}

\subsection{Validation of fine-tuned models}

\begin{figure}
  \centering
  \includegraphics[width=1.0\columnwidth]{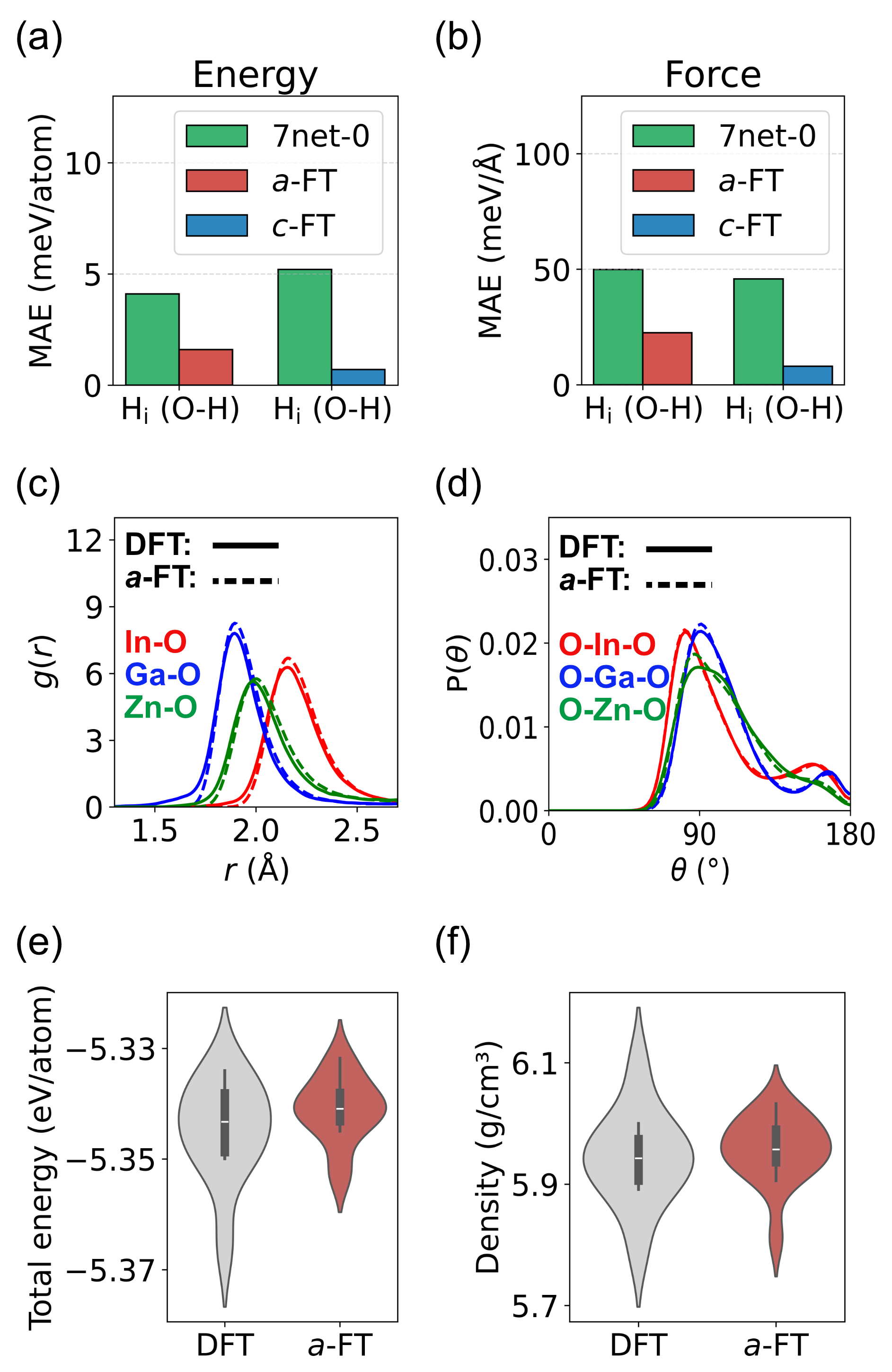}
  \caption{Evaluation of prediction performance of fine-tuned MLIP models. (a) Energy MAEs and (b) force MAEs for systems with H$_\mathrm{i}$ forming a \ce{O-H} bond. (c) Radial distribution function ($g(r)$), (d) angular distribution function ($\rm{P}(\theta)$), (e) total energy per atom, and (f) gravimetric density of pristine $a$-IGZO structures. In (e) and (f), white dots indicate the median and black bars denote the interquartile range. }
  \label{fig:a-FT_summary}
\end{figure}

We first validate the accuracy of the two fine-tuned MLIP models—$a$-FT and $c$-FT. Both models demonstrate improved predictability and robustness relative to DFT results, as evidenced by reduced mean absolute errors (MAEs) and higher $R^2$ values across all test sets, including pristine structures and diverse hydrogen configurations (Figs.~S3–S6). Among the test cases, we closely examine systems containing an interstitial hydrogen impurity forming an \ce{O-H} bond (H$_\mathrm{i}$), which is the most thermodynamically favorable hydrogen defect in IGZO (Figs.~\ref{fig:a-FT_summary}a and b). For comparison, we also present the MAEs of the original 7net-0 model. The fine-tuned $a$-FT MLIP achieves significantly low MAEs—less than 2 meV/atom for energy and less than 23 meV/Å for force. These errors are more than two times lower than those obtained with the original 7net-0 model. A comparable level of improvement is also observed for the $c$-FT MLIP. 
In addition to predictive accuracy, MD simulations based on the $a$-FT model accurately reproduces the distribution of metal–oxygen (\ce{$M$-O}) bond lengths in $a$-IGZO as obtained from FPMD simulations as shown in the radial distribution functions (RDFs) in Fig.~\ref{fig:a-FT_summary}c. The In–O, Ga–O, and Zn–O bond lengths estimated from the corresponding RDF peak positions are in good agreement with previous calculations and experiments~\cite{mo_hosono_2008, mo_cho_2009, mq_kang_2014}. The $a$-FT model also produces angle distributions of metal-centered local motifs in excellent agreement with DFT results (Fig.~\ref{fig:a-FT_summary}d). Beyond local atomic structures, the energies and gravimetric densities of $a$-IGZO configurations obtained by the $a$-FT MLIP also show favorable agreement with DFT calculations, as illustrated in Figs.~\ref{fig:a-FT_summary}e and f, respectively. 

We also confirm the improved predictive accuracy of the fine-tuned models for various types of hydrogen defects, including substitutional hydrogen occupying an oxygen site (H$_\mathrm{O}$), another well-established hydrogen configuration in IGZO systems~\cite{hi_ho_nbis_noh_2013, mh_bang_2017, defects_meux_2018}, as shown in Fig.~S7. Notably, the force prediction accuracy for H$_\mathrm{O}$ configurations—where interactions between hydrogen and neighboring cations are inadequately captured by 7net-0—is significantly enhanced through fine-tuning, as demonstrated in Fig.~S8.

\begin{figure}
  \centering
  \includegraphics[width=1.0\columnwidth]{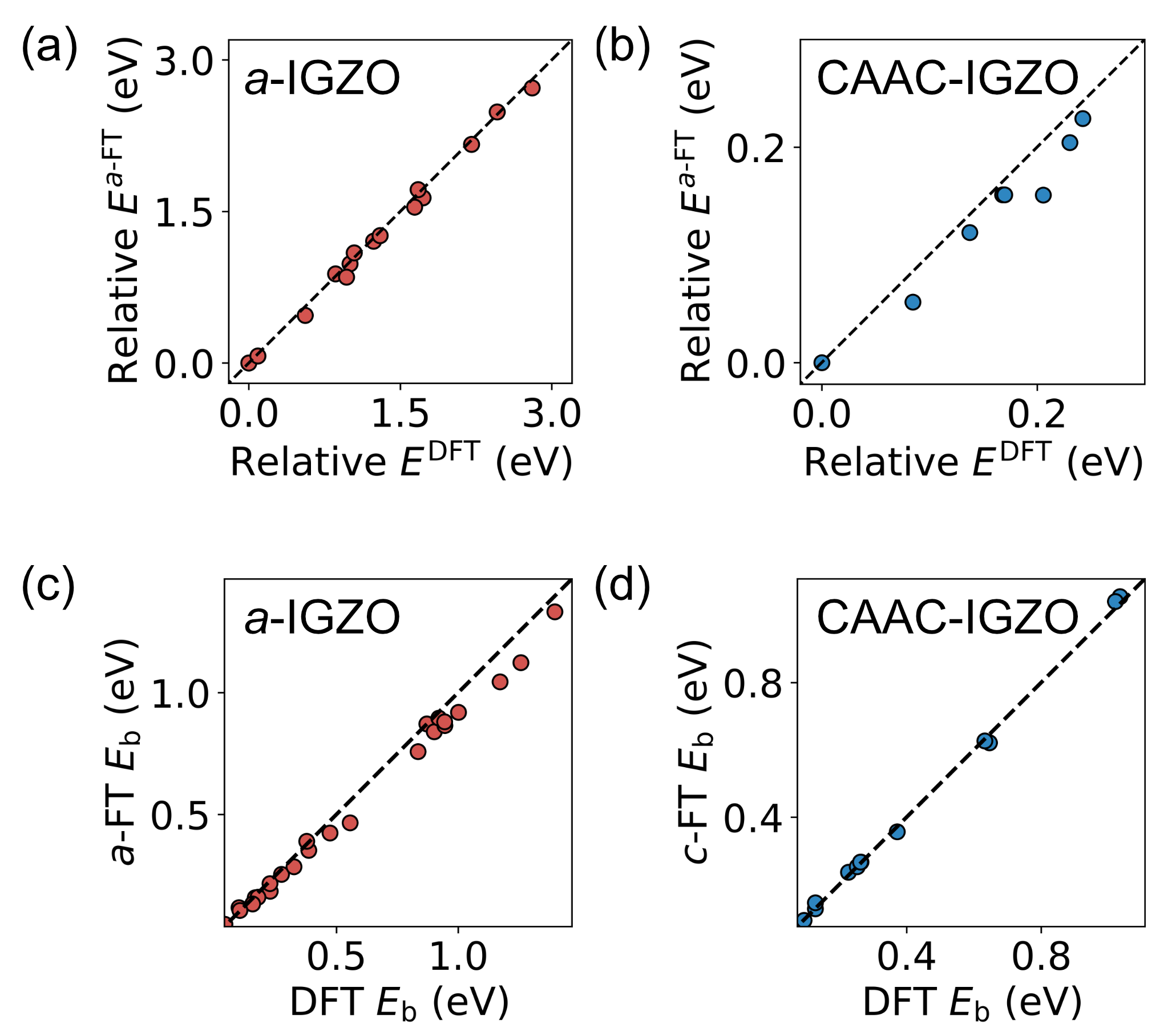}
  \caption{Validation of fine-tuned MLIP models for relative energies of IGZO with various H$_{\rm{i}}$ configurations and migration barriers ($E_{\rm{b}}$s). Relative energies in (a) $a$-IGZO (15 configurations) and (b) CAAC-IGZO (5 configurations). Migration barriers of (c) $a$-IGZO (24 paths) and (d) CAAC-IGZO (12 paths).}
  \label{fig:e_summary}
\end{figure}

Next, we compare the relative energies of various H$_\mathrm{i}$ configurations in $a$-IGZO and CAAC-IGZO systems between DFT and fine-tuned MLIPs. Because hydrogen migrates between different oxygen sites, accurately capturing the energy ordering among H$_i$ configurations is essential for reliably modeling hydrogen dynamics. As shown in Figs.~\ref{fig:e_summary}a and b, the fine-tuned MLIPs successfully reproduce the energy hierarchy for both systems, demonstrating strong linear correlation with DFT results. We further assess the energy ordering of various hydrogen defect types in diverse atomic environments, including oxygen-deficient $a$-IGZO (Fig.~S9), which confirms the excellent predictive accuracy of the $a$-FT model.

We further evaluate whether the fine-tuned MLIP models can accurately reproduce hydrogen migration barriers ($E_{\rm{b}}$) in both $a$- and CAAC-IGZO systems. Long-range hydrogen diffusion in IGZO predominantly proceeds via a combination of hopping and flipping mechanisms, which will be discussed in detail below. To assess model accuracy for these diffusion processes, we select representative hopping and flipping events in both $a$- and CAAC-IGZO and calculate the corresponding migration barriers using both DFT and MLIP-based nudged elastic band (NEB) methods. As shown in Figs.~\ref{fig:e_summary}c and d, the $a$-FT and $c$-FT models successfully reproduce the DFT-calculated migration barriers for their respective systems, demonstrating the high accuracy of the fine-tuned MLIP models in capturing hydrogen diffusion kinetics.

\subsection{H diffusion mechanism}

We perform MD simulations using the fine-tuned MLIP models to investigate hydrogen diffusion in both $a$-IGZO and CAAC-IGZO systems. To ensure a sufficient number of diffusion events and meaningful MSD data within accessible simulation timescales, which is essential for reliably determining hydrogen diffusivity, the simulations are conducted at elevated temperatures above 650 K.

\subsubsection{Amorphous IGZO}

\begin{figure*}
  \centering
  \includegraphics[width=0.8\textwidth]{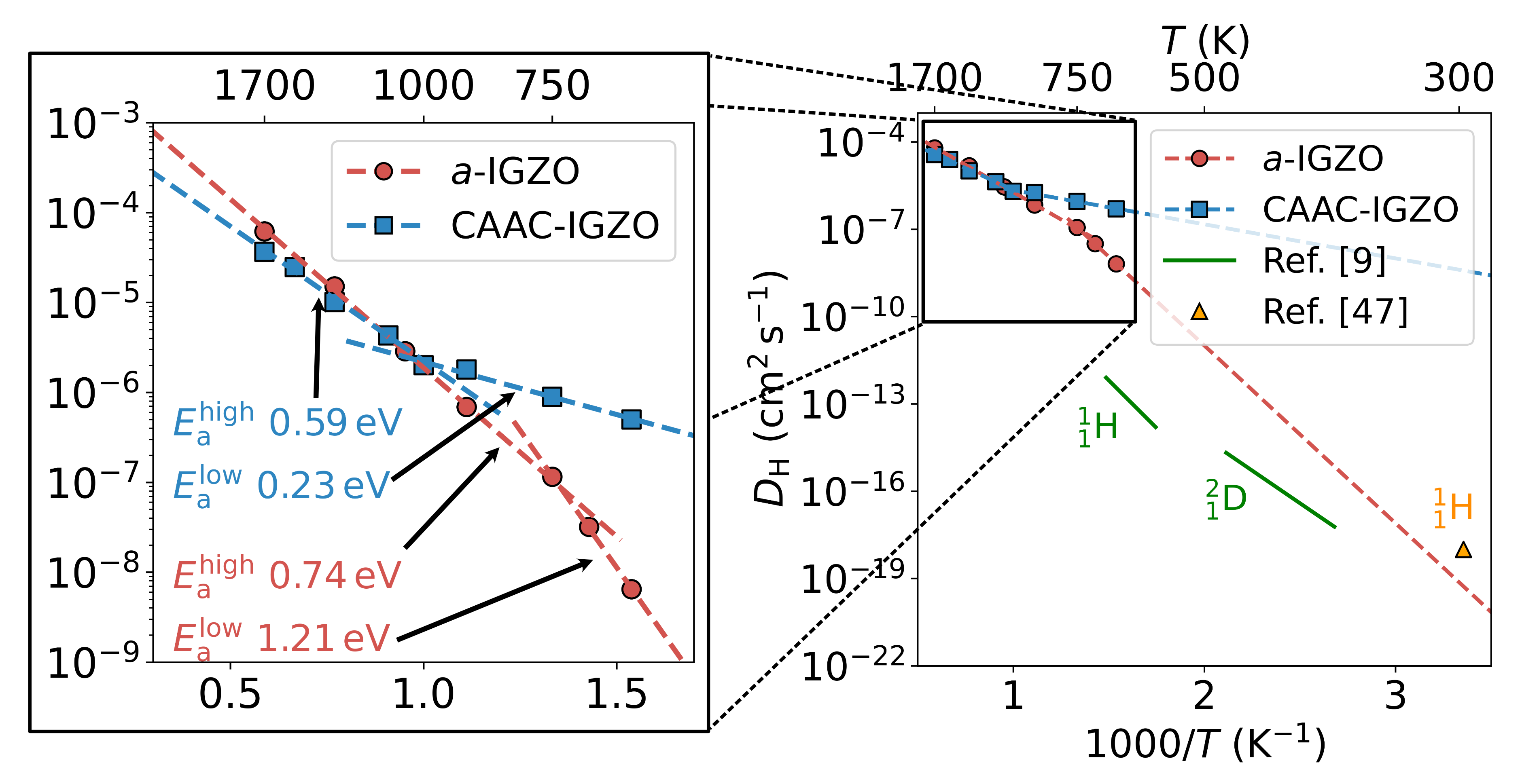}
    \caption{\small{Arrhenius plot of H diffusivity $D_{\rm{H}}$ as a function of $1000/T$, where $T$ is the temperature, in $a$- and CAAC-IGZO. Both systems exhibit two distinct linear regions with different activation energies. $^1_1\rm{H}$ and $^2_1\rm{D}$ indicate Hydrogen and Deuterium, respectively.}}
  \label{fig:h_diffusion_aIGZO}
\end{figure*}

Hydrogen diffusion in $a$-IGZO is investigated over a wide temperature range from 650 to 1700 K. The Arrhenius plots of hydrogen diffusivity are shown in Fig.~\ref{fig:h_diffusion_aIGZO}. Two distinct linear regions with different activation energies are observed, intersecting near 750 K. The low-temperature trend (below 750 K) extrapolated down to 300 K yields diffusivity values that are reasonably consistent with existing experimental data~\cite{h_diffusivity_nomura_2013,h_diffusivity_chung_2015}, considering the associated measurement uncertainties.

\begin{figure*}
  \centering
  \includegraphics[width=0.7\textwidth]{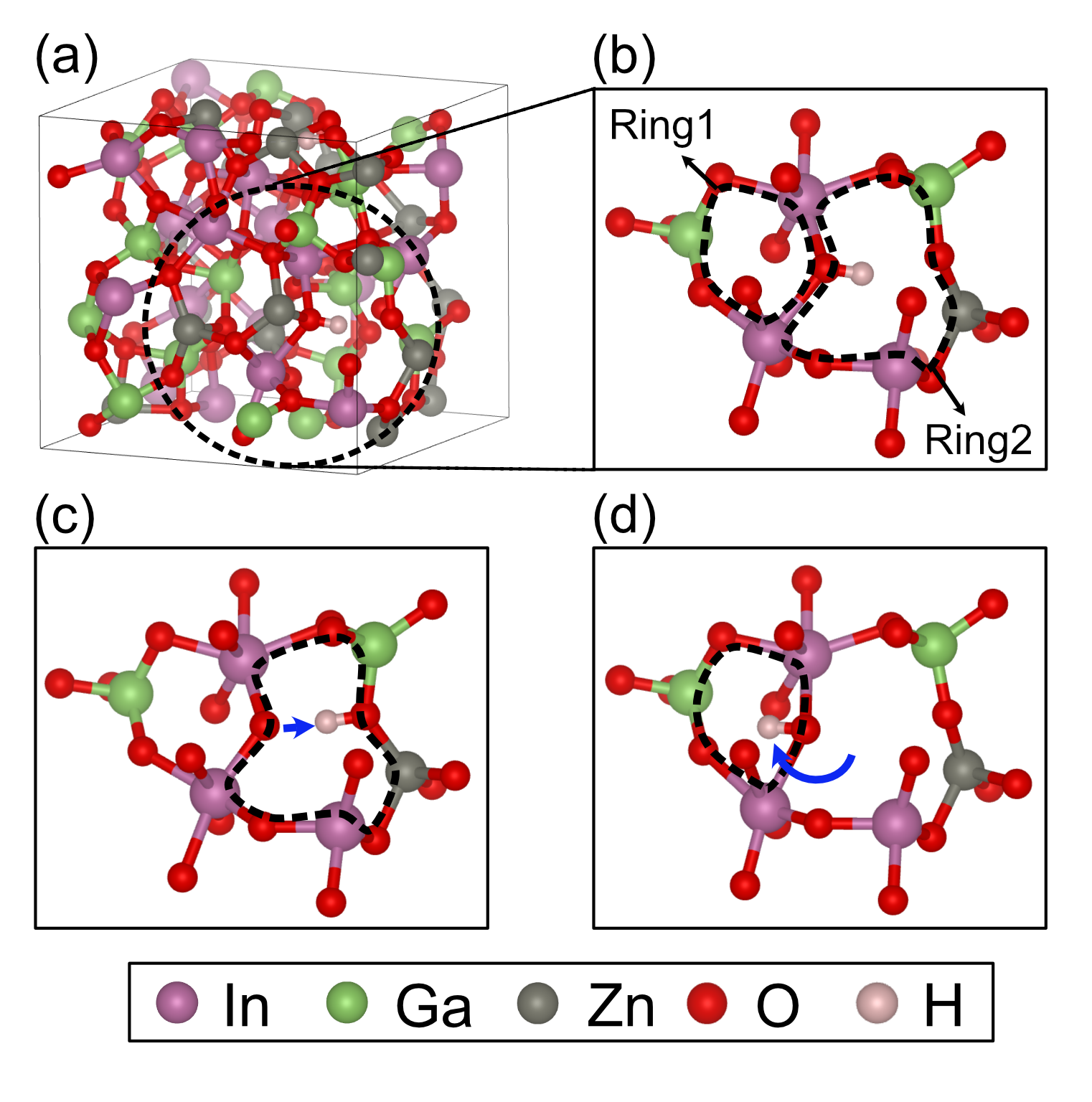}
\caption{ Diffusion processes in $a$-IGZO. (a) $a$-IGZO model, (b) ring configurations, (c) intra-ring hopping (IRH), and (d) ring-to-ring flipping (RTRF). 
}
  \label{fig:a_igzo}
\end{figure*}

We examine hydrogen diffusion processes in $a$-IGZO in the low-temperature regime, based on an analysis of MD trajectories. As shown in Figs.~\ref{fig:a_igzo}a and b, the $a$-IGZO structure features metal-oxygen ring networks, and the introduced hydrogen atoms primarily form \ce{O-H} bonds. In this situation, hydrogen diffusion occurs mainly through two processes: intra-ring hopping (IRH) and ring-to-ring  flipping (RTRF). IRH refers to the hopping of a hydrogen atom between two neighboring oxygen sites within the same ring, typically between sites facing each other (Fig.~\ref{fig:a_igzo}c). Since this process is confined within a single ring, long-range hydrogen diffusion requires an additional process to exit the ring structure. RTRF enables such an escape through a large-angle reorientation of the \ce{O-H} bond (Fig.~\ref{fig:a_igzo}d). Following a successful RTRF transition, the hydrogen atom can further migrate to a new O site within the newly accessed ring via IRH. Repeated sequences of RTRF followed by IRH allow hydrogen to propagate across multiple ring structures, thereby enabling long-range diffusion.  

\begin{figure*}[p]
  \centering
  \includegraphics[width=1.0\textwidth]{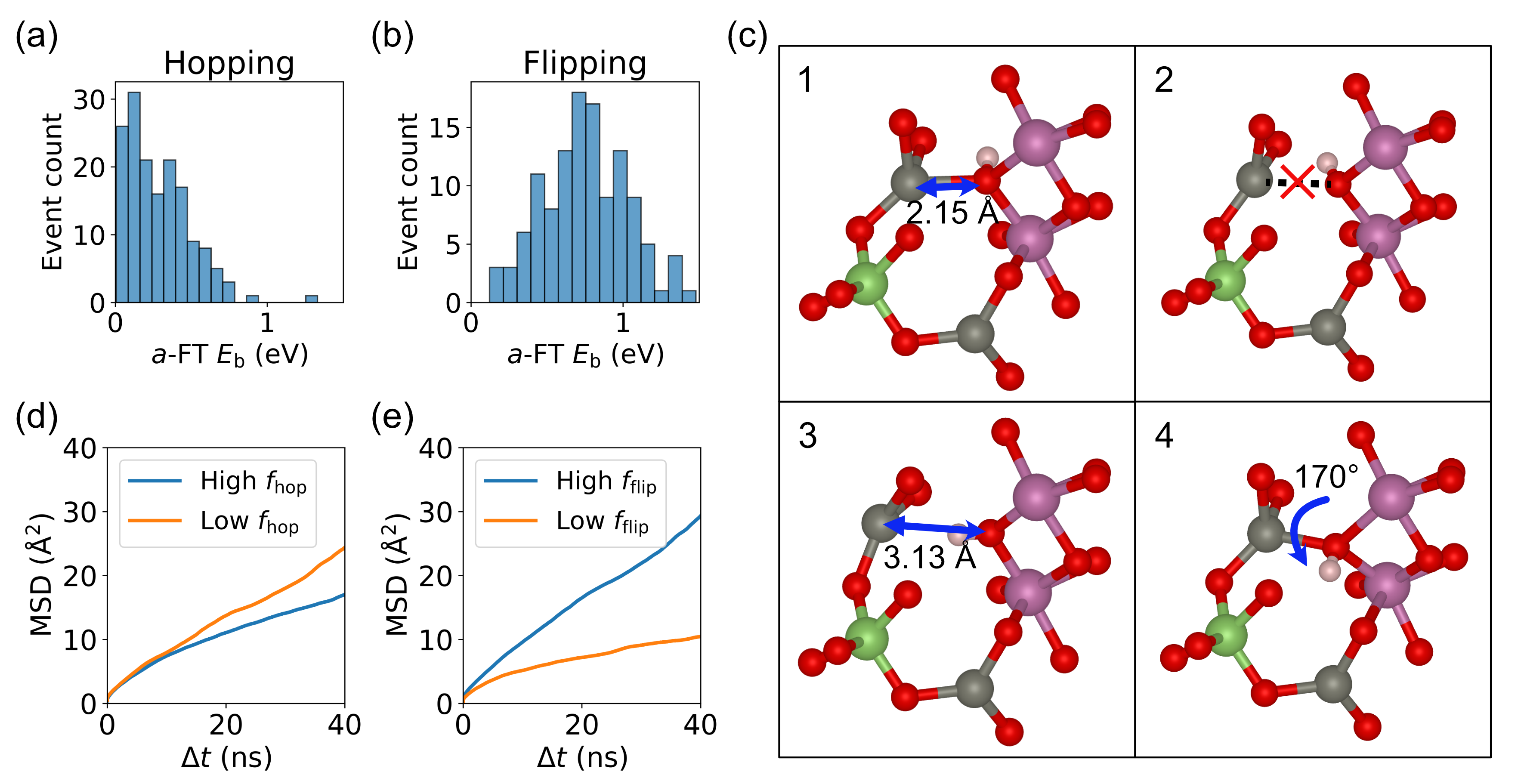}
  \caption{\small{
 Energy barriers for (a) hydrogen hopping (159 samples) and (b) hydrogen flipping (121 samples). (c) NEB images of a representative flipping process, illustrating the sequential steps of \ce{$M$-O} bond breaking, \ce{O-H} bond rotation, and \ce{$M$-O} bond reconstruction. Contributions of high- and low-frequency transition groups for (d) hopping and (e) flipping processes to hydrogen MSD at 650 K. High and low $f_{\rm{hop}}$ values are 847~$\mathrm{ns}^{-1}$ and 122~$\mathrm{ns}^{-1}$, respectively, and High and low $f_{\rm{flip}}$ values are 99~$\mathrm{ns}^{-1}$ and 14~$\mathrm{ns}^{-1}$, respectively. 
}}
  \label{fig:msd_event}
\end{figure*}

We calculate the energy barriers for IRH and RTRF using NEB calculations. To account for statistical variations inherent in amorphous systems, we sample multiple diffusion pathways by MLIP NEB—159 for IRH and 121 for RTRF. For IRH, most energy barriers are found to be below 0.5~eV, with an average of approximately 0.29~eV (Fig.~\ref{fig:msd_event}a). In contrast, RTRF exhibits higher energy barriers, with an average of 0.76~eV (Fig.~\ref{fig:msd_event}b), making it a much slower process and the rate-determining step for long-range hydrogen diffusion. The relatively high barrier of RTRF is attributed to the temporary breaking of an \ce{$M$-O} bond during the transition, which is subsequently reformed (Fig.~\ref{fig:msd_event}c).

\clearpage

As discussed above, RTRF processes are likely to determine the overall rate of long-range hydrogen diffusion, and thus, govern its diffusivity. The critical role of RTRF is further validated by quantifying the contributions of hopping and flipping to hydrogen transport. To this end, we analyze 48 H trajectories from a 60-ns MD simulation at 650 K. For each trajectory, the total number of hopping and flipping events is counted and converted into event frequencies ($f_{\rm{hop}}$ for hopping and $f_{\rm{flip}}$ for flipping). Here, flipping is considered to occur when a H atom rotates more than $90^\circ$. Fig.~\ref{fig:msd_event}d illustrates averaged MSD as a function of $\Delta t$ for two groups of hydrogen atoms: the top 50\% with the highest $f_{\rm{hop}}$s, and the remaining 50\%. The two groups exhibit similar MSDs, indicating a weak correlation between MSD and $f_{\rm{hop}}$. This suggests that, despite occurring more frequently than flipping (with average frequencies of $f_{\rm{hop}}=485~\mathrm{ns}^{-1}$ and $f_{\rm{flip}}=56~\mathrm{ns}^{-1}$), hopping processes alone are not critical in determining the overall diffusivity. This is because hopping typically takes place within a single ring and often occurs in a back-and-forth manner, contributing little to net displacement. In contrast, Fig.~\ref{fig:msd_event}e shows the MSD for two groups of hydrogen atoms classified by $f_{\rm{flip}}$, and evidently, hydrogen atoms with higher $f_{\rm{flip}}$s display significantly larger MSDs. This observation highlights the importance of RTRF processes in enabling long-range hydrogen diffusion.

Thus far, we have focused on hydrogen diffusion below 750 K. In this low-temperature regime, the \ce{$M$-O} structural framework remains relatively rigid, with minimal displacement of metal and oxygen atoms over the simulation time. In contrast, at elevated temperatures above 750 K, the increased thermal energy and high structural flexibility of amorphous phase allow for greater atomic displacements of metal and oxygen atoms (Fig.~S12). Under these conditions, the instantaneous breaking of \ce{$M$-O} bonds occurs more frequently, facilitating \ce{O-H} flipping processes that are inaccessible at lower temperatures. Furthermore, without breaking the \ce{O-H} bond, oxygen atoms can migrate while carrying the \ce{O-H} unit through the disordered atomic network, thereby expanding the available diffusion pathways for hydrogen.
These effects collectively lead to a reduction in the effective activation energy for hydrogen diffusion at high temperatures.

\subsubsection{$c$-axis aligned crystalline IGZO}

\begin{figure*}
  \centering
  \includegraphics[width=1\textwidth]{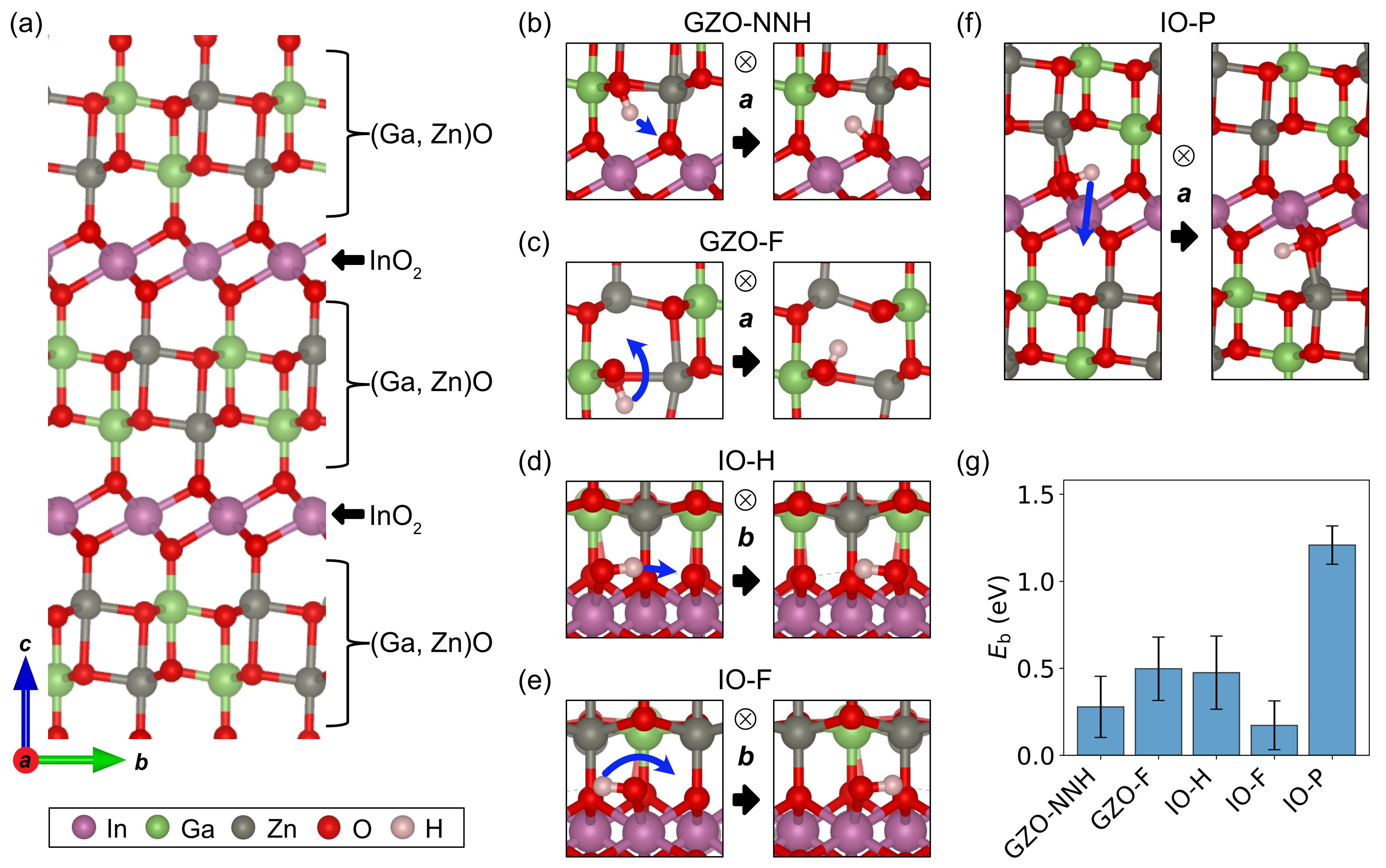}
  \caption{(a) CAAC-IGZO crystal structure. Identified H migration processes: (b) \ce{Ga/Zn-O} nearest neighbor hopping (GZO-NNH), (c) \ce{Ga/Zn-O} \textit{c}-axis flipping (GZO-F), (d) \ce{In-O} \textit{ab}-plane hopping (IO-H), (e) \ce{In-O} \textit{ab}-plane flipping (IO-F), and (f) \ce{In-O} \textit{c}-axis penetrating (IO-P). (g) Averaged migration barrier and standard deviation from $c$-FT MLIP calculations for each process. The number of samples are 112 for GZO-NNH, 112 for GZO-F, 106 for IO-H, 24 for IO-F, and 32 for IO-P.}
  \label{fig:caac_igzo}
\end{figure*}

\begin{figure}
  \centering
  \includegraphics[width=0.8\columnwidth]{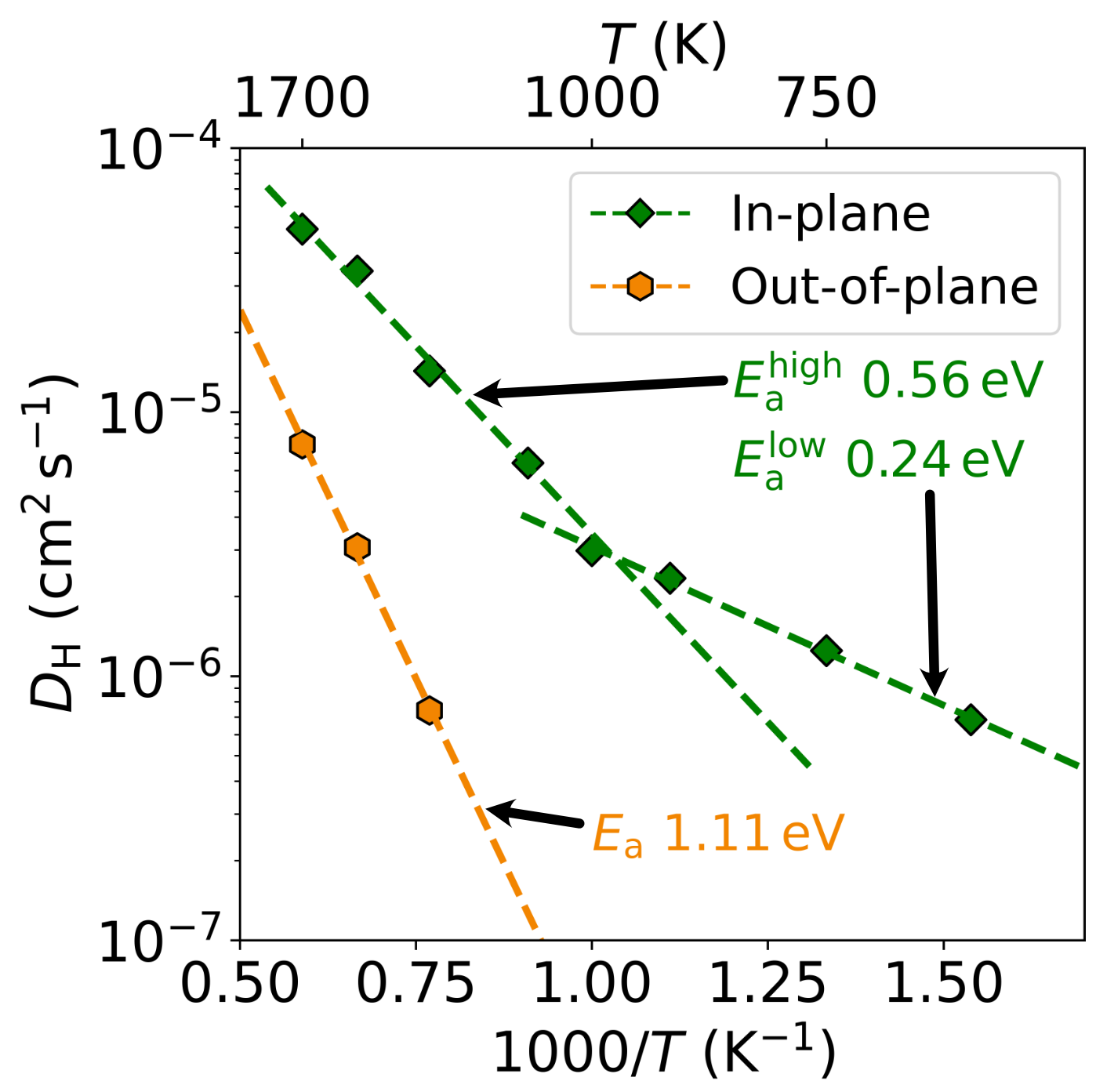}
  \caption{
  Arrhenius plot of in-plane and out-of-plane $D_{\rm{H}}$ as a function of $1000/T$, where $T$ is the temperature, in CAAC-IGZO. It shows highly anisotropic diffusion process of CAAC-IGZO.
  \label{fig:Fig8}
  }
\end{figure}

Fig.~\ref{fig:caac_igzo}a shows the crystal structure of CAAC-IGZO composed of \ce{In-O} and \ce{Ga/Zn-O} layers alternating along the $c$-axis. Within the \ce{Ga/Zn-O} layers, Ga and Zn randomly occupy the same crystallographic sites. Hydrogen atoms incorporated into CAAC-IGZO primarily form O–H bonds and preferentially reside within the \ce{Ga/Zn-O} layers or at the boundary between \ce{Ga/Zn-O} and \ce{In-O} layers, whereas incorporation within the \ce{In-O} layers is energetically unfavorable~\cite{caac_defects_omura_2009, caac_chen_2011}. Hydrogen diffusion in CAAC-IGZO is investigated at temperatures ranging from 650 to 1700 K, and the resulting Arrhenius plot reveals two distinct linear regimes, as shown in Fig.~\ref{fig:h_diffusion_aIGZO}. At low-temperature regime, including room temperature, CAAC-IGZO exhibits a higher diffusivity than $a$-IGZO. However, unlike $a$-IGZO, the diffusion process in CAAC-IGZO is highly anisotropic, with \textit{ab}-plane (in-plane) direction diffusion significantly exceeding that along the \textit{c}-axis (Fig.~\ref{fig:Fig8}). 

To uncover the origin of the anisotropic hydrogen transport, we analyze hydrogen trajectories from MLIP-MD simulations and identify five representative diffusion processes, as illustrated in Figs.~\ref{fig:caac_igzo}b–f: (i)  Ga/Zn–O nearest-neighbor hopping (GZO-NNH), (ii)  Ga/Zn–O \textit{c}-axis flipping (GZO-F), (iii) In–O \textit{ab}-plane hopping (IO-H), (iv) In–O \textit{ab}-plane flipping (IO-F), and (v) In–O \textit{c}-axis penetrating (IO-P). The GZO-NNH process involves hydrogen hopping between oxygen atoms coordinated to a common Ga or Zn atom, exhibiting low energy barriers averaging around 0.28 eV. IO-F processes also exhibit low migration barriers—about 0.17 eV on average—and facilitate hydrogen escape from confined hopping sites through reorientation of a \ce{O-H} bond. The combination of these low-barrier processes enables long-range hydrogen diffusion along the \textit{ab}-plane through a zigzag hopping pathway. On the other hand, the GZO-F and IO-H processes can also contribute to in-plane hydrogen diffusion. However, they exhibit relatively higher energy barriers—approximately 0.50~eV for GZO-F and 0.48~eV for IO-H on average—which makes their contribution weaker than those of the low-barrier GZO-NNH and IO-F processes. 
At low temperatures, hydrogen diffusion is particularly dominated by the GZO-NNH and IO-F pathways, whose individual energy barriers closely correspond to the low-temperature activation energy ($E_\mathrm{a}$) for in-plane diffusion (Figs.~\ref{fig:Fig8}). However, as the temperature increases, higher-barrier processes such as GZO-F and IO-H also become activated. This increases the effective activation energy and simultaneously accelerates hydrogen diffusion along the \textit{ab}-plane. This is the main reason for the transition of the diffusivity curve at $~\sim$1000 K in Figs.~\ref{fig:h_diffusion_aIGZO} and ~\ref{fig:Fig8}.

Notably, we observe that hydrogen diffusion across the \ce{In-O} layer along the \textit{c}-axis (out-of-plane)—a pathway critical for vertical long-range transport—is significantly suppressed (Fig.~\ref{fig:caac_igzo}f). NEB calculations reveal that the energy barrier for IO-P pathway is exceptionally high, approaching 1.2 eV (Fig.~\ref{fig:caac_igzo}g), closely matching $E_\mathrm{a}=1.11~\rm{eV}$ obtained from the Arrhenius plot of out-of-plane diffusion (Fig.~\ref{fig:Fig8}). Such high barriers explain the negligible  diffusion events along the $c$ direction in MD simulations, particularly at low-temperature region.

\subsection{Kinetic stability of $M$-H bonds}
\begin{figure*}
  \centering
  \includegraphics[width=0.9\textwidth]{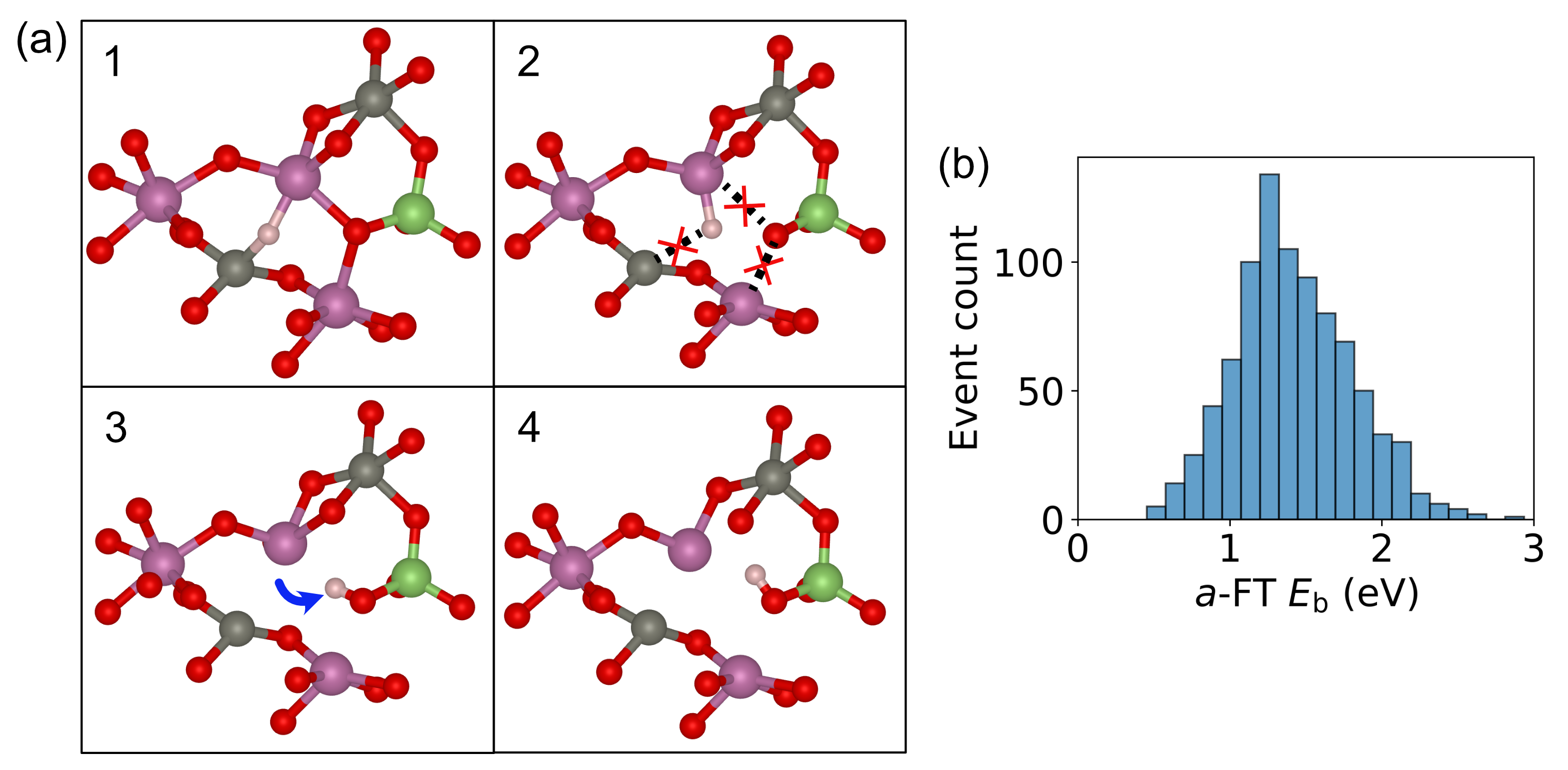}
  \caption{\small{
  (a) NEB images of a representative $\mathrm{H_O}$ to $\mathrm{H_i}$ transition, illustrating the simultaneous breaking of \ce{$M$-H} and \ce{$M$-O} bonds involved in the process. (b) Distribution of hydrogen migration barriers obtained from NEB calculations over 868 distinct pathways.}}
  \label{fig:mh2oh}
\end{figure*}

Although hydrogen is known to preferentially form an \ce{O-H} bond in $a$-IGZO systems, it can also bond with metal atoms by either substituting an oxygen site or occupying an interstitial site surrounded by metal atoms~\cite{h_bistability_kang_2015, mh_bang_2017}. These metal-coordinated hydrogen configurations, which involve $M$–H bonding, may be introduced during thin-film fabrication processes. Due to their significant kinetic stability, such configurations are suggested to persist and remain immobile at typical device operating temperatures (300–400~K). To assess their kinetic stability, we evaluate the energy barriers for hydrogen escape from a metal-coordinated H$_\mathrm{O}$ site to an interstitial H$_\mathrm{i}$ site forming an \ce{O-H} bond in $a$-IGZO (Fig.~\ref{fig:mh2oh}a). A total of 868 MLIP NEB calculations are performed for distinct diffusion pathways, yielding a relatively high average barrier of 1.42 eV (Fig.~\ref{fig:mh2oh}b). This barrier leads to a negligible Arrhenius factor of approximately $10^{-21}$ at 350~K, indicating that such transitions are effectively suppressed under typical operating conditions. The high energy barriers are attributed to the simultaneous breaking of \ce{$M$-H} and \ce{$M$-O} bonds during the transition, as illustrated in Fig.~\ref{fig:mh2oh}a.

\subsection{Effects of hydrogen diffusion on device performance}
As mentioned in the Introduction, the negative shift in $V_\mathrm{th}$ observed in oxide TFTs under NB(I)S conditions has been attributed to the diffusion of positively charged hydrogen—present in the form of \ce{O-H} bonds in the $a$-IGZO channel—toward the gate insulator interface. This hypothesis implicitly assumes that hydrogen is sufficiently mobile to reach the interface within the stress duration; however, this has not been conclusively demonstrated. Based on the Arrhenius relation shown in Fig.~\ref{fig:h_diffusion_aIGZO}, the hydrogen diffusivity is estimated to be approximately $D_\mathrm{H} = 5.9 \times 10^{-17}~\mathrm{cm}^2/\mathrm{s}$ at 350~K. This corresponds to a diffusion length, given by $\sqrt{2D_{\rm{H}}t}$, ranging from approximately 1.1 to 11~nm for time scales between $10^2$ and $10^4$ seconds. These diffusion lengths are comparable to the typical channel thickness, approximately 20–50~nm in display devices~\cite{review_ide_2019} and below 20~nm in memory devices~\cite{memory_nm_belmonte_2020,memory_nm_tang_2024}, suggesting that hydrogen migration is likely to contribute to the observed $V_\mathrm{th}$ shift in experiments involving stress durations exceeding $10^2$ seconds. Furthermore, since hydrogen diffusion is a thermally activated process—i.e., its diffusivity increases with temperature—its impact is expected to be more pronounced at elevated temperatures. Indeed, previous experiments reported that device instability is exacerbated with increasing temperature and longer bias application times~\cite{temp_nbs_chang_2014}. As a note, under NB(I)S conditions, hydrogen diffusion may be accelerated by the electric field. However, this effect is likely confined to the vicinity of the interface, as the external bias is predominantly applied to the gate insulator and the channel region immediately adjacent to it~\cite{nbs_li_2019}. Therefore, most hydrogen ions in the channel, which are located more than several nanometers away from the interface, are expected to have diffusion lengths comparable to our estimates.

In planar oxide TFTs used for display applications, hydrogen diffusion along the channel length direction during the TFT fabrication also plays a critical role in determining the carrier density—and thus the $V_\mathrm{th}$. According to a previous work~\cite{h_instability_chen_2020}, micrometer-scale $a$-IGZO channels in top-gate TFTs exhibit lower $V_\mathrm{th}$ values (i.e., higher carrier densities) as the channel length decreases. This experimental observation has been attributed to more substantial and uniform hydrogen incorporation in shorter channels during the plasma-enhanced chemical vapor deposition (PECVD) of the SiO$_2$ interlayer dielectric (ILD) on top of them. This explanation is based on the assumption that hydrogen generated during PECVD diffuses laterally into the oxide channel from the sidewalls in contact with the source and drain electrodes, eventually reaching the channel center in short channels. Given the ILD deposition duration of 30 minutes, the estimated diffusion lengths based on our calculated diffusivity are approximately 4.6~\textmu m and 36.5~\textmu m at process temperatures of 260 $^\circ\mathrm{C}$ and 360 $^\circ\mathrm{C}$, respectively. While the actual amount of hydrogen incorporated into the channel depends on the diffusion process through the ILD and across the ILD/channel interface, our results clearly indicate that channels shorter than 10~\textmu m can be significantly affected by hydrogen incorporation during ILD deposition—consistent with experimental observations. Therefore, to achieve a $V_\mathrm{th}$ near 0 V, precise control over hydrogen incorporation during the ILD deposition is essential and should be tailored to a given channel length.  

On the other hand, compared to $a$-IGZO, CAAC-IGZO has been experimentally reported to exhibit significantly greater resistance to electrical instabilities under NB(I)S conditions~\cite{caac_park_2015, caac_a_nc_kang_2021}. Notably, as in $a$-IGZO, hydrogen concentrations exceeding 10$^{19}~$cm$^{-3}$ can be incorporated into CAAC-IGZO thin films during channel deposition\cite{h_conc_caac_ono_2020}. One key factor contributing to the superior bias stability of CAAC-IGZO is the suppressed hydrogen diffusion along the $c$-axis, as demonstrated in our simulations. When we resolve the hydrogen diffusivity by crystallographic direction, the out-of-plane (i.e., $c$-axis) component is estimated to be only $1.7\times 10^{-18}~\mathrm{cm}^2/\mathrm{s}$ at 350 K, corresponding to a diffusion length about one order of magnitude shorter than in $a$-IGZO. Consequently, hydrogen in the CAAC-IGZO channel is far less likely to reach the gate insulator interface under NB(I)S conditions compared to hydrogen in $a$-IGZO, contributing to the enhanced reliability of CAAC-based devices. Note that the smaller $c$-axis diffusivity in CAAC-IGZO compared to $a$-IGZO, despite its lower $E_{\rm{a}}$, arises from its lower diffusion pre-factor $D_0$ (Tables~S4 and S7). In amorphous systems with a more flexible atomic network than their crystalline counterparts, transitions are often associated with larger entropy changes, arising from more extended collective atomic motions that require the activation of a larger number of phonon modes (Fig.~S13)~\cite{zener_naundorf_1999,mn_yelon_2006}.

\section{Conclusions}

In this study, we conducted a comprehensive theoretical investigation of hydrogen diffusion in $a$- and CAAC-IGZO using MLIP-MD simulations. By fine-tuning the universal pretrained SevenNet-0 model, we developed phase-specific MLIP models that accurately capture the structural and dynamical behaviors of hydrogen in both disordered and crystalline environments. Our results reveal that in $a$-IGZO, long-range hydrogen transport is facilitated by a combination of intra-ring hopping and ring-to-ring flipping mechanisms. In contrast, hydrogen diffusion in CAAC-IGZO is highly anisotropic, dominated by in-plane motion, with vertical ($c$-axis) diffusion strongly suppressed.

Extrapolated diffusivities at operating temperatures (300–400~K) indicate that hydrogen in $a$-IGZO can reach the channel/insulator interface within typical stress durations, contributing to $V_{\rm{th}}$ instability under NB(I)S conditions. In contrast, the suppressed vertical diffusion in CAAC-IGZO suggests a reduced impact on bias-induced degradation. Furthermore, our study highlights that hydrogen incorporation during PECVD of the SiO$_2$ ILD can lead to enhanced $n$-type doping in shorter channels, thereby influencing $V_{\rm{th}}$ and contributing to the observed channel-length-dependent behavior in oxide TFTs. Overall, our study offers a detailed microscopic understanding of hydrogen transport mechanisms in IGZO-based semiconductors, establishing a foundation for designing oxide TFTs with high stability and performance.

\section*{Author contributions}
H. Cho: conceptualization, methodology, investigation, validation, formal analysis, data curation, and writing – original draft;
M. Moon and J. Kim: investigation and methodology support;
E. Koh: project coordination, communication, and resource management;
H.-D. Kim: project administration and resources;
R. Kim and G. Park: project support and resources;
S. Han: supervision and project involvement; and
Y. Kang: conceptualization, supervision, methodology, data interpretation, and writing – review \& editing;

\section*{Conflicts of interest}

There are no conflicts to declare.

\section*{Data availability}
Supplementary data and additional information are provided in the Electronic Supplementary Information (ESI) of this article. Training data for the MLIPs will be available upon request, except for the open database. The SevenNet code used in this study is available in the GitHub repository: https://www.github.com/MDIL-SNU/SevenNet.

\section*{Acknowledgements}
This research was supported by Samsung Display Company Ltd and the National Research Foundation (NRF) funded by the Korean government (MSIT) (No. RS-2024-00407840). The computations were carried out at the Korea Institute of Science and Technology Information (KISTI) National Supercomputing Center (KSC-2025-CRE-0110) and at the Center for Advanced Computations (CAC) at Korea Institute for Advanced Study (KIAS). 

%%%END OF MAIN TEXT%%%


\balance


%%%REFERENCES%%%
\bibliography{reference} %You need to replace "rsc" on this line with the name of your .bib file
\bibliographystyle{reference} %the RSC's .bst file
\end{document}